
%
%
%
\hsize=6.5truein
\hoffset=0truein
\vsize=8.9truein
\voffset=0truein
\font\twelverm=cmr10 scaled 1200    \font\twelvei=cmmi10 scaled 1200
\font\twelvesy=cmsy10 scaled 1200   \font\twelveex=cmex10 scaled 1200
\font\twelvebf=cmbx10 scaled 1200   \font\twelvesl=cmsl10 scaled 1200
\font\twelvett=cmtt10 scaled 1200   \font\twelveit=cmti10 scaled 1200
\skewchar\twelvei='177   \skewchar\twelvesy='60
\def\twelvepoint{\normalbaselineskip=12.4pt
  \abovedisplayskip 12.4pt plus 3pt minus 9pt
  \belowdisplayskip 12.4pt plus 3pt minus 9pt
  \abovedisplayshortskip 0pt plus 3pt
  \belowdisplayshortskip 7.2pt plus 3pt minus 4pt
  \smallskipamount=3.6pt plus1.2pt minus1.2pt
  \medskipamount=7.2pt plus2.4pt minus2.4pt
  \bigskipamount=14.4pt plus4.8pt minus4.8pt
  \def\rm{\fam0\twelverm}          \def\it{\fam\itfam\twelveit}%
  \def\sl{\fam\slfam\twelvesl}     \def\bf{\fam\bffam\twelvebf}%
  \def\mit{\fam 1}                 \def\cal{\fam 2}%
  \def\tt{\twelvett}
  \textfont0=\twelverm   \scriptfont0=\tenrm   \scriptscriptfont0=\sevenrm
  \textfont1=\twelvei    \scriptfont1=\teni    \scriptscriptfont1=\seveni
  \textfont2=\twelvesy   \scriptfont2=\tensy   \scriptscriptfont2=\sevensy
  \textfont3=\twelveex   \scriptfont3=\twelveex  \scriptscriptfont3=\twelveex
  \textfont\itfam=\twelveit
  \textfont\slfam=\twelvesl
  \textfont\bffam=\twelvebf \scriptfont\bffam=\tenbf
  \scriptscriptfont\bffam=\sevenbf
  \normalbaselines\rm}
\font\titlerm=cmr10 scaled\magstep3
\font\titleit=cmti10 scaled\magstep3

\def\titlefonts{\def\rm{\titlerm}
                \def\it{\titleit} \rm}
\font\twelvesc=cmcsc10 scaled 1200
\def\beginlinemode{\endmode
  \begingroup\parskip=0pt \obeylines\def\\{\par}\def\endmode{\par\endgroup}}
\def\beginparmode{\endmode
  \begingroup \def\endmode{\par\endgroup}}
\let\endmode=\par
{\obeylines\gdef\
{}}
\def\singlespace{\baselineskip=\normalbaselineskip}
\def\oneandahalfspace{\baselineskip=\normalbaselineskip
  \multiply\baselineskip by 3 \divide\baselineskip by 2}
\def\doublespace{\baselineskip=\normalbaselineskip \multiply\baselineskip by 2}

\newcount\firstpageno
\firstpageno=2
\footline={\ifnum\pageno<\firstpageno{\hfil}\else{\hfil\twelverm\folio\hfil}\fi}
\let\rawfootnote=\footnote		
\def\footnote#1#2{{\rm\singlespace\parindent=0pt\rawfootnote{#1}{#2}}}
\def\raggedcenter{\leftskip=4em plus 12em \rightskip=\leftskip
  \parindent=0pt \parfillskip=0pt \spaceskip=.3333em \xspaceskip=.5em
  \pretolerance=9999 \tolerance=9999
  \hyphenpenalty=9999 \exhyphenpenalty=9999 }
\parskip=\medskipamount
\twelvepoint		
\doublespace		
\overfullrule=0pt	
\def\preprintno#1{
 \rightline{\rm #1}}	
\def\title                      
  {\null\vskip 3pt plus 0.2fill
   \beginlinemode \doublespace \raggedcenter \titlefonts}
\def\author                     
  {\vskip 3pt plus 0.2fill \beginlinemode
   \singlespace \raggedcenter\twelvesc}
\def\affil			
  {\vskip 3pt plus 0.03fill \beginlinemode
   \oneandahalfspace \it \centerline}
\def\abstract			
  {\vskip 3pt plus 0.3fill \beginparmode
   \doublespace \narrower ABSTRACT: }
\def\endtitlepage		
  {\endpage			
   \body}
\def\body			
  {\beginparmode}		
\def\head#1{			
  \filbreak\vskip 0.5truein	
  {\immediate\write16{#1}
   \raggedcenter \bf\uppercase{#1}\par}
   \nobreak\vskip 0.25truein\nobreak}
\def\subhead#1{			
  \vskip 0.25truein		
  {\raggedcenter \bf{#1}\par}
   \nobreak\vskip 0.25truein\nobreak}

\def\references
  {\head{REFERENCES}
   \frenchspacing \parindent=0pt \leftskip=0.8truecm \rightskip=0truecm
   \parskip=4pt plus 2pt \everypar{\hangindent=\parindent}}
\def\pr{\journal Phys. Rev.}

\def\cmp{\journal Comm. Math. Phys.}
\def\np{\journal Nucl. Phys.}
\def\pl{\journal Phys. Lett.}

\def\zphys{\journal Z. Phys.}

\def\ptp{\journal Prog. Theor. Phys.}

\def\endreferences{\body}
\def\figurecaptions		
  {\endpage
   \beginparmode
   \head{Figure Captions}
}

\def\endpage			
  {\vfill\eject}
\def\endpaper			
  {\endmode\vfill\supereject}
\def\endit
  {\endpaper\end}

\def\frac#1#2{{\textstyle{#1 \over #2}}}
\def\ref#1{ref. #1}
\def\Ref#1{Ref. #1}

\def\eg{{\it e.g.}}

\def\ie{{\it i.e.}}

\def\etal{{\it et al.}}

\def\sla{\raise.15ex\hbox{$/$}\kern-.57em}
\def\leaderfill{\leaders\hbox to 1em{\hss.\hss}\hfill}
\def\twiddle{\lower.9ex\rlap{$\kern-.1em\scriptstyle\sim$}}
\def\bigtwiddle{\lower1.ex\rlap{$\sim$}}
\def\gtwid{\mathrel{\raise.3ex\hbox{$>$\kern-.75em\lower1ex\hbox{$\sim$}}}}
\def\ltwid{\mathrel{\raise.3ex\hbox{$<$\kern-.75em\lower1ex\hbox{$\sim$}}}}
\def\square{\kern1pt\vbox{\hrule height 1.2pt\hbox{\vrule width 1.2pt\hskip 3pt
   \vbox{\vskip 6pt}\hskip 3pt\vrule width 0.6pt}\hrule height 0.6pt}\kern1pt}

%
%
%
\def\refstylenp{		
  \gdef\refto##1{~[##1]}				
  \gdef\r##1{~[##1]}	         			
  \gdef\refis##1{\indent\hbox to 0pt{\hss[##1]~}}     	
  \gdef\citerange##1##2##3{\cite{##1}--\setbox0=\hbox{\cite{##2}}\cite{##3}}
  \gdef\journal##1, ##2, ##3,                           
    ##4,{{\sl##1} {\bf ##2} (##3) ##4}}
\def\refstylezphys{		
  \gdef\refto##1{~[##1]}				
  \gdef\r##1{~[##1]}	         			
  \gdef\refis##1{\indent\hbox to 0pt{\hss[##1]~}}     	
  \gdef\citerange##1##2##3{\cite{##1}--\setbox0=\hbox{\cite{##2}}\cite{##3}}
  \gdef\journal##1, ##2, ##3,                           
    ##4,{{\sl##1} {\bf ##2}, ##4 (##3)}}
\def\refstylepr{		
  \gdef\refto##1{~[##1]}		
  \gdef\r##1{~[##1]}		        
  \gdef\refis##1{\indent\hbox to 0pt{\hss[##1]~}}	
  \gdef\citerange##1##2##3{\cite{##1}--\setbox0=\hbox{\cite{##2}}\cite{##3}}
  \gdef\journal##1, ##2, ##3,                           
    ##4,{{\sl##1} {\bf ##2}, ##4 (##3)}}
\def\refstyleijmp{		
  \gdef\refto##1{$^{##1}$}				
  \gdef\r##1{$^{##1}$}	         			
  \gdef\refis##1{\indent\hbox to 0pt{\hss##1.~}}     	
  \gdef\citerange##1##2##3{\cite{##1}--\setbox0=\hbox{\cite{##2}}\cite{##3}}
  \gdef\journal##1, ##2, ##3,                           
    ##4,{{\sl##1} {\bf ##2} (##3) ##4}}
\def\(#1){(\call{#1})}
\def\call#1{{#1}}
\def\smu
{Department of Physics, Southern Methodist University, Dallas, TX 75275}

\def\ths{\thinspace}

\def\e{\epsilon}
\def\g{\gamma}
\def\d{\delta}

\def\s{\sigma}
\def\l{\lambda}

\def\vp{\varphi}
\def\del{\partial}
\def\ha{{1\over 2}}

\def\ra{\rightarrow}

\def\ub#1{\underline{#1}}
\def\ob#1{\overline{#1}}

\def\kd3{\delta^{(3)}}
\def\sym#1,#2{\Biggl\{ {#1}{1\over i\del^+} {#2} \Biggr\}_{\rm sym} }
\def\ssym#1,#2{\Biggl\{ {#1}{1\over (i\del^+)^2} {#2} \Biggr\}_{\rm sym} }
\refstylenp
\catcode`@=11
\newcount\tagnumber\tagnumber=0
\immediate\newwrite\eqnfile
\newif\if@qnfile\@qnfilefalse
\def\write@qn#1{}
\def\writenew@qn#1{}
\def\w@rnwrite#1{\write@qn{#1}\message{#1}}
\def\@rrwrite#1{\write@qn{#1}\errmessage{#1}}
\def\taghead#1{\gdef\t@ghead{#1}\global\tagnumber=0}
\def\t@ghead{}
\expandafter\def\csname @qnnum-3\endcsname
  {{\t@ghead\advance\tagnumber by -3\relax\number\tagnumber}}
\expandafter\def\csname @qnnum-2\endcsname
  {{\t@ghead\advance\tagnumber by -2\relax\number\tagnumber}}
\expandafter\def\csname @qnnum-1\endcsname
  {{\t@ghead\advance\tagnumber by -1\relax\number\tagnumber}}
\expandafter\def\csname @qnnum0\endcsname
  {\t@ghead\number\tagnumber}
\expandafter\def\csname @qnnum+1\endcsname
  {{\t@ghead\advance\tagnumber by 1\relax\number\tagnumber}}
\expandafter\def\csname @qnnum+2\endcsname
  {{\t@ghead\advance\tagnumber by 2\relax\number\tagnumber}}
\expandafter\def\csname @qnnum+3\endcsname
  {{\t@ghead\advance\tagnumber by 3\relax\number\tagnumber}}
\def\equationfile{%
  \@qnfiletrue\immediate\openout\eqnfile=\jobname.eqn%
  \def\write@qn##1{\if@qnfile\immediate\write\eqnfile{##1}\fi}
  \def\writenew@qn##1{\if@qnfile\immediate\write\eqnfile
    {\noexpand\tag{##1} = (\t@ghead\number\tagnumber)}\fi}
}
\def\callall#1{\xdef#1##1{#1{\noexpand\call{##1}}}}
\def\call#1{\each@rg\callr@nge{#1}}
\def\each@rg#1#2{{\let\thecsname=#1\expandafter\first@rg#2,\end,}}
\def\first@rg#1,{\thecsname{#1}\apply@rg}
\def\apply@rg#1,{\ifx\end#1\let\next=\relax%
\else,\thecsname{#1}\let\next=\apply@rg\fi\next}
\def\callr@nge#1{\calldor@nge#1-\end-}
\def\callr@ngeat#1\end-{#1}
\def\calldor@nge#1-#2-{\ifx\end#2\@qneatspace#1 %
  \else\calll@@p{#1}{#2}\callr@ngeat\fi}
\def\calll@@p#1#2{\ifnum#1>#2{\@rrwrite{Equation range #1-#2\space is bad.}
\errhelp{If you call a series of equations by the notation M-N, then M and
N must be integers, and N must be greater than or equal to M.}}\else%
 {\count0=#1\count1=#2\advance\count1
by1\relax\expandafter\@qncall\the\count0,%
  \loop\advance\count0 by1\relax%
    \ifnum\count0<\count1,\expandafter\@qncall\the\count0,%
  \repeat}\fi}
\def\@qneatspace#1#2 {\@qncall#1#2,}
\def\@qncall#1,{\ifunc@lled{#1}{\def\next{#1}\ifx\next\empty\else
  \w@rnwrite{Equation number \noexpand\(>>#1<<) has not been defined yet.}
  >>#1<<\fi}\else\csname @qnnum#1\endcsname\fi}
\let\eqnono=\eqno
\def\eqno(#1){\tag#1}
\def\tag#1$${\eqnono(\displayt@g#1 )$$}
\def\aligntag#1\endaligntag
  $${\gdef\tag##1\\{&(##1 )\cr}\eqalignno{#1\\}$$
  \gdef\tag##1$${\eqnono(\displayt@g##1 )$$}}

\def\eqalignno#1{\displ@y \tabskip\centering
  \halign to\displaywidth{\hfil$\displaystyle{##}$\tabskip\z@skip
    &$\displaystyle{{}##}$\hfil\tabskip\centering
    &\llap{$\displayt@gpar##$}\tabskip\z@skip\crcr
    #1\crcr}}
\def\displayt@gpar(#1){(\displayt@g#1 )}
\def\displayt@g#1 {\rm\ifunc@lled{#1}\global\advance\tagnumber by1
        {\def\next{#1}\ifx\next\empty\else\expandafter
        \xdef\csname @qnnum#1\endcsname{\t@ghead\number\tagnumber}\fi}%
  \writenew@qn{#1}\t@ghead\number\tagnumber\else
        {\edef\next{\t@ghead\number\tagnumber}%
        \expandafter\ifx\csname @qnnum#1\endcsname\next\else
        \w@rnwrite{Equation \noexpand\tag{#1} is a duplicate number.}\fi}%
  \csname @qnnum#1\endcsname\fi}
\def\ifunc@lled#1{\expandafter\ifx\csname @qnnum#1\endcsname\relax}
\let\@qnend=\end\gdef\end{\if@qnfile
\immediate\write16{Equation numbers written on []\jobname.EQN.}\fi\@qnend}
\newcount\r@fcount \r@fcount=0
\newcount\r@fcurr
\immediate\newwrite\reffile
\newif\ifr@ffile\r@ffilefalse
\def\w@rnwrite#1{\ifr@ffile\immediate\write\reffile{#1}\fi\message{#1}}
\def\writer@f#1>>{}
\def\referencefile{
  \r@ffiletrue\immediate\openout\reffile=\jobname.ref%
  \def\writer@f##1>>{\ifr@ffile\immediate\write\reffile%
    {\noexpand\refis{##1} = \csname r@fnum##1\endcsname = %
     \expandafter\expandafter\expandafter\strip@t\expandafter%
     \meaning\csname r@ftext\csname r@fnum##1\endcsname\endcsname}\fi}%
  \def\strip@t##1>>{}}

\def\citeall#1{\xdef#1##1{#1{\noexpand\cite{##1}}}}
\def\cite#1{\each@rg\citer@nge{#1}}	
\def\each@rg#1#2{{\let\thecsname=#1\expandafter\first@rg#2,\end,}}
\def\first@rg#1,{\thecsname{#1}\apply@rg}	
\def\apply@rg#1,{\ifx\end#1\let\next=\relax
\else,\thecsname{#1}\let\next=\apply@rg\fi\next}
\def\citer@nge#1{\citedor@nge#1-\end-}	
\def\citer@ngeat#1\end-{#1}
\def\citedor@nge#1-#2-{\ifx\end#2\r@featspace#1 
  \else\citel@@p{#1}{#2}\citer@ngeat\fi}	
\def\citel@@p#1#2{\ifnum#1>#2{\errmessage{Reference range #1-#2\space is bad.}%
    \errhelp{If you cite a series of references by the notation M-N, then M and
    N must be integers, and N must be greater than or equal to M.}}\else%
 {\count0=#1\count1=#2\advance\count1
by1\relax\expandafter\r@fcite\the\count0,%
  \loop\advance\count0 by1\relax
    \ifnum\count0<\count1,\expandafter\r@fcite\the\count0,%
  \repeat}\fi}
\def\r@featspace#1#2 {\r@fcite#1#2,}	
\def\r@fcite#1,{\ifuncit@d{#1}
    \newr@f{#1}%
    \expandafter\gdef\csname r@ftext\number\r@fcount\endcsname%
                     {\message{Reference #1 to be supplied.}%
                      \writer@f#1>>#1 to be supplied.\par}%
 \fi%
 \csname r@fnum#1\endcsname}
\def\ifuncit@d#1{\expandafter\ifx\csname r@fnum#1\endcsname\relax}%
\def\newr@f#1{\global\advance\r@fcount by1%
    \expandafter\xdef\csname r@fnum#1\endcsname{\number\r@fcount}}
\let\r@fis=\refis			
\def\refis#1#2#3\par{\ifuncit@d{#1}
   \newr@f{#1}%
   \w@rnwrite{Reference #1=\number\r@fcount\space is not cited up to now.}\fi%
  \expandafter\gdef\csname r@ftext\csname r@fnum#1\endcsname\endcsname%
  {\writer@f#1>>#2#3\par}}
\def\ignoreuncited{
   \def\refis##1##2##3\par{\ifuncit@d{##1}%
     \else\expandafter\gdef\csname r@ftext\csname
r@fnum##1\endcsname\endcsname%
     {\writer@f##1>>##2##3\par}\fi}}
\def\r@ferr{\endreferences\errmessage{I was expecting to see
\noexpand\endreferences before now;  I have inserted it here.}}
\let\r@ferences=\references
\def\references{\r@ferences\def\endmode{\r@ferr\par\endgroup}}
\let\endr@ferences=\endreferences
\def\endreferences{\r@fcurr=0
  {\loop\ifnum\r@fcurr<\r@fcount
    \advance\r@fcurr by 1\relax\expandafter\r@fis\expandafter{\number\r@fcurr}%
    \csname r@ftext\number\r@fcurr\endcsname%
  \repeat}\gdef\r@ferr{}\endr@ferences}
\let\r@fend=\endpaper\gdef\endpaper{\ifr@ffile
\immediate\write16{Cross References written on []\jobname.REF.}\fi\r@fend}
\catcode`@=12
\citeall\refto		
\citeall\ref		%
\citeall\Ref		%
\citeall\r		%
\ignoreuncited
\def\eg{{\it e.g.}}

\singlespace
\preprintno{SMUHEP/93--20}
\doublespace

\title Light-Cone Quantization of Gauge Fields
\author Gary McCartor$^a$ and David G. Robertson$^b$
\affil\smu

\abstract\
Light-cone quantization of gauge field theory is considered.  With a
careful treatment of the relevant degrees of freedom and where they
must be initialized, the results obtained in equal-time quantization
are recovered, in particular the Mandelstam-Leibbrandt form of the
gauge field propagator.  Some aspects of the ``discretized''
light-cone quantization of gauge fields are discussed.

\footnote{}{$^a$Email: mccartor@mail.physics.smu.edu}
\footnote{}{$^b$Present address: Department of Physics, The Ohio State
University, Columbus, OH 43210.  Email: dgr@pacific.mps.ohio-state.edu}

\endtitlepage
\oneandahalfspace

\taghead{1.}
\subhead{1. Introduction}

When setting up a light-cone quantized version of a gauge theory it is
most natural to choose the light-cone gauge: $n\cdot A=0,$ with
$n^2=0.$ This choice is particularly convenient when $n\cdot x$ is the
evolution parameter of the system.  (We shall define $x^\pm\equiv
x^0\pm x^3,$ and take $x^+$ to be the evolution parameter.  Thus we
take $n_\mu=(1,0,0,1).$) In this case Gauss' law appears to be a
constraint relation, which can be solved for $A^-$ in terms of the
transverse degrees of freedom.  Thus one can eliminate all unphysical
components of the gauge field at the level of the Hamiltonian.  The
Fadeev-Popov sector decouples, so that unitarity is manifest.  And
finally, the light-cone gauge, like other algebraic noncovariant
gauges, is free of the Gribov ambiguity.

However, implementation of this gauge is not entirely straightforward,
at least in perturbation theory.  The problem is easily stated: the
naive gauge field propagator
$$
D^{ab}_{\mu\nu}(k) ={i\d^{ab}\over k^2+i\e}
\Bigl[-g_{\mu\nu}+{n_\mu k_\nu + n_\nu k_\mu \over n\cdot k}\Bigr]
					\eqno(naiveprop)
$$
is singular at $n\cdot k = k^0+k^3=0$. (In a Hamiltonian framework this
appears as singularities in certain matrix elements of the Hamiltonian.)
The issue is how this object is to be given a meaning.

Various suggestions have been made in this regard.  The simplest
approach is to interpret the singularity as a Cauchy principal
value\r{ks70}.  This can easily be seen to be wrong, however, on
physical grounds.  The basic problem is that the spurious poles lie in
different quadrants of the complex $k^0$-plane than do the usual
Feynman poles.  Thus positive (negative) energy quanta do not
necessarily propagate into the forward (backward) light cone.
Practically speaking, what happens is that a Wick rotation is
impossible without crossing a spurious pole, and extra contributions
are generated.  These extra terms are responsible for the
well-documented failure of calculations performed in this way to agree
with calculations performed in, \eg, covariant gauges\r{bassettobook}.

A more successful prescription for the spurious singularity is that
suggested independently by Mandelstam\r{mandelstam83} and
Leibbrandt\r{leibbrandt84} (ML).  They define (in Leibbrandt's
notation)
$$
{1\over [k^+]_{\rm ML}}\equiv {k^-\over k^+k^-+i\e} ,
					\eqno(whatlmis)
$$
so that the pole in $k^+$ is shifted above or below the real axis
depending on the sign of $k^-$.  With this definition, the spurious
poles are distributed in the same way as the Feynman poles; thus no
poles are encountered in performing a Wick rotation, and extra terms
are not generated.  Many explicit calculations have been performed
using the ML prescription, and all give sensible results, in
agreement, where comparison is possible, with covariant-gauge results.

Bassetto, \etal\ have given a derivation of the ML propagator in the
framework of equal-time quantization\r{bdls85}, and have further shown
that gauge theories defined in this way are renormalizable\r{bds87}
(although nonlocal counterterms are necessary to render off-shell
Green functions finite).  A central feature of their construction is
that they do not reduce down to the physical (transverse) degrees of
freedom.  A longitudinal gauge degree of freedom is retained, and a
corresponding ghost field.  In a way familiar from covariant-gauge
quantization, selection of a physical subspace results in the recovery
of a positive-semidefinite metric and Poincar\'e invariance.  More
concretely, one considers the Lagrangian
$$
{\cal L}=-{1\over4}F_{\mu\nu}^a F^{\mu\nu a} - \lambda^a n_\mu A^{\mu a} ,
					\eqno(lagrangian)
$$
with $\l^a$ a Lagrange multiplier field whose equation of motion
enforces the gauge condition $A^{+a}=0.$  In addition to $\l^a,$ one
keeps the other three components of the gauge field as degrees of
freedom.  Equal-time quantization of this Lagrangian leads
unambiguously to the ML form of the propagator.  It is an open---and
interesting---question whether a formulation exists in which only
physical degrees of freedom are retained and the gluon propagator is
well defined.

Light-cone quantization of the Lagrangian \(lagrangian), however, is
slightly mysterious.  The puzzle has to do with the field $\l^a$; it
satisfies
$$
\del_-\l^a=0 ,
					\eqno(leom)
$$
which is an equation of constraint on the light-cone initial-value
surface $x^+=0.$  Thus if we impose the natural boundary condition that
$\l^a$ vanish at $x^-=\pm\infty$, we obtain $\l^a=0$ everywhere, and a
formalism that cannot be equivalent to the equal-time quantized
version of \(lagrangian).  In particular, we are led to a principal
value prescription (or equivalent) for the spurious pole.

The purpose of the present paper is to show how a careful light-cone
quantization of the Lagrangian \(lagrangian) in fact leads to
precisely the results obtained by Bassetto, \etal, and in particular
to the ML form of the gauge field propagator.  The main point we shall
make is that certain of the degrees of freedom---including the field
$\l^a$---must be initialized along a surface of constant $x^-$, in a
way familiar from the treatment of massless fields in two spacetime
dimensions, for example.

We shall begin in Section 2 by identifying the relevant degrees of
freedom and how they must be initialized.  We then discuss the
construction of the dynamical operators $P^\pm$ in the free theory and
the determination of the commutation relations to be satisfied by the
fields.  The free theory is trivially solved, and we compute the
propagator.  Finally, we show how the physical subspace is to be
selected, and the Poincar\'e invariance of the theory recovered.
Section 3 contains some remarks on the problem of quantizing a gauge
field with a periodicity condition imposed along a characteristic
surface---the case of ``discretized'' light-cone quantization.  Some
concluding remarks are presented in Section 4.

Another promising subtraction procedure, inspired by the ML
prescription and realizable in the context of light-cone perturbation
theory, has recently been proposed\r{bassetto91,soperheidelberg,ls93}.
Its consistency has yet to be explored in detail, however.

\taghead{2.}
\subhead{2. Light-Cone Quantization of the Lagrangian \(lagrangian)}

In this section we show how a careful light-cone quantization of the
Lagrangian \(lagrangian) leads to the same results obtained in
equal-time quantization\r{bdls85}.  We shall begin by considering the
free theory ($g=0$) and, to avoid spurious complexities, only a single
gauge field.  That is, we study the problem of QED with no fermions.

Our first task is to use the equations of motion to identify the
degrees of freedom of the system, that is, those data that are
necessary to allow the most general solution to the equations of
motion throughout all of spacetime.  These data will correspond to
independent operators in the quantum field theory, whose commutation
relations are determined by demanding that the Heisenberg equations of
motion correctly reproduce the classical field equations, and that the
Poincar\'e algebra be correctly realized.  The most straightforward
way to proceed is simply to imagine using the field equations to
evolve the fields from some assumed classical initial data.  If all of
spacetime can be filled out in this way, then the assumed data is
sufficient.  One then tries to uncover redundancies, in order to
arrive at the minimal data that allows a completely general solution.
These minimal data then correspond to operators which we must include
when we go over to the quantum field theory.

The procedure is most easily explained by example.  Let us consider
the case of a free, massless Dirac field in two spacetime dimensions.
As is well known, the two-spinor $\psi$ can be broken up into two
one-spinors $\psi_\pm\equiv \ha\g^0\g^\pm\psi$, corresponding (in two
dimensions) to chirality eigenstates.  In terms of these fields the
Dirac equation separates into
$$
\del_+\psi_+=0
					\eqno(dirac2p)
$$
$$
\del_-\psi_-=0.
					\eqno(dirac2m)
$$
Now, the solution of eqs. \(dirac2p)--\(dirac2m) is trivial: $\psi_+$
($\psi_-$) can be any arbitrary function of $x^-$ ($x^+$).  In the
language of evolving from initial data, we might say that we specify
$\psi_+$ along $x^+=0$ and $\psi_-$ along $x^-=0.$ We then use eqs.
\(dirac2p)--\(dirac2m) to evolve the fields off of these surfaces;
although in this case the evolution is particularly simple.  Each
field is simply constant in its ``evolution parameter.'' The initial
data is clearly sufficient to determine $\psi_\pm$ everywhere, and it
is also clear that it corresponds to the most general solution of the
free, massless Dirac equation in two dimensions.  When we build the
quantum theory, then, we must include {\it both} a $\psi_+$ field,
satisfying an equal-$x^+$ commutator, and a $\psi_-$ field satisfying
an equal-$x^-$ commutator.  Throwing away some of these degrees of
freedom---setting $\psi_-$ to zero, for example---results in a theory
which is not isomorphic to the equal-time theory of a massless Dirac
particle\r{mccartor88}.  In particular, if we leave out the $\psi_-$
degrees of freedom then the theory will contain only right-moving
particles.

One lesson of this example is that it is not always possible when
quantizing on characteristic surfaces to have a strictly
``Hamiltonian'' framework, with all the dynamical variables evolving
from a single initial-value surface.  In general, degrees of freedom
initialized on other characteristics may need to be included.  (The
unusual feature is not so much that the initial-value surface has a
kink in it, as that some variables are initialized on one part while
others are initialized on another part.)  Particular subsets of the
full solution space may be selected by imposing boundary conditions on
the fields, but the resulting theory will not generally represent all
of the desired physics.  In the case of the (1+1)-dimensional massless
fermion, for example, we could demand that $\psi_-$ vanish at
$x^-=\infty.$ This would then force $\psi_-=0$ everywhere, leading to
a {\it different} theory.  Some physics---in this case, the physics of
left-moving particles---has been excluded.

Let us now consider the gauge theory.  The equations of motion
obtained from the Abelian version of \(lagrangian) can be cast in the
compact form
$$
\del_\mu F^{\mu\nu}=n^\nu\l .
					\eqno(compact)
$$
Writing these out for the different values of $\nu$, in terms of
light-cone variables and with $A^+$ set to zero, we have
$$
(\del_-)^2A^-+\del_-\del_iA^i=0
					\eqno(eom1)
$$
$$
2\del_+\del_-A^--\del_\perp^2A^--2\del_+\del_iA^i=2\l
					\eqno(eom2)
$$
$$
(4\del_+\del_- - \del_\perp^2)A^i + \del_i(\del_-A^-+\del_jA^j)=0 .
					\eqno(eom3)
$$
Note also that, from eq. \(compact), the field $\l$ satisfies
$$
n\cdot\del\l=\del_-\l=0.
					\eqno(lambdaeom)
$$

Our first observation is that the solution of eq. \(lambdaeom) is
simply that $\l$ be an arbitrary function of $(x^+,x_\perp)$.  {\it
Thus $\l$ is analogous to the $\psi_-$ field in the example above; it
must be initialized along a surface of constant $x^-.$} Different
choices for $\l$ give different solutions to the field equations, so
in the quantum theory we must have degrees of freedom corresponding to
the freedom to specify $\l.$ These fields will satisfy an equal-$x^-$
commutation relation with some ``conjugate momentum'' yet to be
determined, and their contribution to conserved charges will be
obtained by integrating suitable components of the charge densities
over their initial-value surface.  Note that one could impose, for
example, the condition that $\l$ vanish at $x^-=\pm\infty,$ leading to
$\l=0$ everywhere.  As in the previous example, however, this results
in the selection of a subspace of the full solution space; the
resulting field theory will not contain all the relevant degrees of
freedom.

Let us now consider eq. \(eom1).  This may be solved to yield
$$
\del_-A^-+\del_iA^i=\vp(x^+,x_\perp)
					\eqno(phi)
$$
where $\vp$ is again arbitrary (note that $\vp$ is just $\del_\mu
A^\mu$ in this gauge).  This is another function which must be
specified to determine the complete solution of the field equations,
and will correspond to degrees of freedom in the quantum theory.
Again, by imposing a condition on $\del_-A^-$ at $x^-=\pm\infty$ we
could remove this freedom, but the resulting field theory would be
incomplete.

The solution of eq. \(phi) apparently involves yet another
undetermined function of $(x^+,x_\perp)$:
$$
A^-=-{1\over\del_-}\del_iA^i + x^-\vp + \g(x^+,x_\perp)
					\eqno(gamma)
$$
(where $(\del_-)^{-1}$ is some particular antiderivative), but we
shall see below that there is a constraint relating $\g,$ $\vp,$ and
$\l,$ so that only two of them are in fact independent.  This is an
example of an apparent freedom in the system which is actually
redundant.

Finally, we must give some initial data for $A^i.$  It is convenient
to define
$$
A^i\equiv T^i + {\del_i\over\del_\perp^2}\vp ,
					\eqno(atdefined)
$$
so that eq. \(eom3) becomes simply
$$
(4\del_+\del_- - \del_\perp^2)T^i=0.
					\eqno(ateom)
$$
The initial data appropriate for a massless scalar field in 3+1
dimensions are the values of the field along $x^+=0.$ The only
subtlety concerns the set of modes with $k_\perp$ and $k^+$
identically zero.  These represent quanta propagating precisely along
the surface $x^+=0,$ so that they can not be initialized there.  They
are somewhat awkward to treat in a continuum formalism, as it is
difficult to properly ``measure'' their contribution.  (Note that they
are of measure zero even relative to the ``zero mode'' fields $\l$ and
$\vp$.)  This is a standard problem in light-cone quantization, which
is conventionally treated by choosing the test functions in which to
smear the fields to vanish at $k^+=0$ \r{ss72}.  Thus they can be
effectively neglected, with the added bonus that the integral operator
$(\del_-)^{-1}$ will be uniquely defined when acting on $T^i$ (it has
no $k^+=0$ mode).

We have now exhausted eqs. \(eom1) and \(eom3), and it is clear that
from these we can determine the solution for $A_\mu$ everywhere in
spacetime.  All that remains is to insure that the solution thus
obtained is consistent with the last equation, \(eom2).  To this end,
we insert the definitions of $\vp$, $\gamma$, and $T^i$ (eqs. \(phi),
\(gamma), and \(atdefined), respectively) into eq.  \(eom2).  Making
use of eq. \(ateom), along with the fact that $(\del_-)^{-1}T^i$ is
uniquely defined, we obtain
$$
-\del_\perp^2\g - 2\del_+\vp = 2\l .
					\eqno(constraint)
$$
Thus the three ``zero mode'' fields $\l$, $\vp$, and $\g$ are not
independent, as promised.  We shall here take $\g$ to be the
determined quantity, and treat $\l$ and $\vp$ as independent fields.

Our final result is that the initial data required to determine the
most general solution to the classical field equations are: $T^i$ on
the traditional light-cone initial-value surface $x^+=0;$ and any two
of $\l,$ $\vp,$ and $\g$ on a surface of constant $x^-.$ (Of course,
these fields are by definition independent of $x^-,$ so it does not
matter exactly where we put them.  For reasons which should become
clear below, we shall imagine them to be initialized along the
boundary ``wings'' at $x^-=\pm\infty$---surface 2 in Fig. 1.)  The
next step is to determine the commutation relations satisfied by these
fields, by constructing the Poincar\'e generators and demanding that
the Heisenberg equations correctly reproduce the field equations
\(eom1)--\(eom3).

The main complicating feature in the computation of conserved charges
is that we must include contributions from the boundary surfaces (Fig.
1).  This actually follows quite generally if we insist that the
charges we construct be identical to those we would obtain in
equal-time quantization\r{mccartor88}.  We have
$$
\del_\mu T^{\mu\nu}=0
					\eqno(consoftmunu)
$$
so that
$$
\oint T^{\mu\nu} d\s_\mu =0 ,
					\eqno(divtheorem)
$$
where the integral is taken over a closed surface.  If this surface is
taken to be that shown in Fig. 1, then it is clear that the integral
over the equal-time surface $t=0$ is equal to that over the light-cone
surface, including the boundary wings.  Thus in
general we must retain contributions coming from the
boundary surfaces to be
assured of obtaining correct results.  In some cases---for example
when only massive fields are present---it may be consistent to discard
the boundary contributions.  In effect, we can assume that massive
fields are sufficiently damped at infinity so that these contributions
are negligible.  When massless fields are present, however, this is
generally not the case.

Let us focus on the generators of translations $P^\pm$.  The (Noether)
energy-momentum tensor derived from the Lagrangian \(lagrangian) is
$$
T^{\mu\nu}_{\rm Noether}=-F^\mu_{~\s}F^{\nu\s}+{1\over4}g^{\mu\nu}F^2
-\l n^\mu A^\nu - \del_\s(F^{\mu\s}A^\nu)
					\eqno(preemt)
$$
As usual, this is neither symmetric nor gauge invariant, but satisfies
$\del_\mu T^{\mu\nu}_{\rm Noether}=0.$ By virtue of the antisymmetry of
$F^{\mu\nu},$ the four-divergence of the last term in eq. \(preemt)
itself vanishes; thus the modified tensor
$$
T^{\mu\nu}=-F^\mu_{~\s}F^{\nu\s}+{1\over4}g^{\mu\nu}F^2
-\l n^\mu A^\nu
					\eqno(emt)
$$
will also give time-independent generators.  We shall here take \(emt)
as the definition of the energy-momentum tensor.

We consider first $P^-$.  We have
$$
P^-=\ha \int_1 dx^-d^2x_\perp T^{+-} + \ha\int_2 dx^+d^2x_\perp
	T^{--} ,
					\eqno(pminus)
$$
where, from eq. \(emt),
$$
T^{+-} = (\del_-A^-)^2 + (\del_iA^j)(\del_iA^j)-(\del_iA^j)(\del_jA^i)
					\eqno(tpm)
$$
$$
T^{--} = 4(\del_+A^i)^2 + (\del_iA^-)^2+4(\del_+A^i)(\del_iA^-)
		-2\l A^- ,
					\eqno(tmm)
$$
and the surfaces 1 and 2 are shown in Fig. 1.
These expressions simplify somewhat when written in terms of $\vp,$
$\l,$ and $T^i$.  Inserting eqs. \(phi) and \(atdefined) into eq.
\(tpm), for example, and integrating by parts freely in the transverse
directions, we find that the first term in eq. \(pminus) can be
written as
$$
\ha\int_1 dx^-d^2x_\perp (\del_iT^j)(\del_iT^j) .
					\eqno(cont1)
$$
Similarly, we may insert the relations
$$
\del_-A^- = -\del_iT^i
					\eqno(dmaminus)
$$
and
$$
A^-= -{1\over\del_-}\del_iT^i -{2\over\del_\perp^2}(\l+\del_+\vp)
					\eqno(aminus)
$$
into eq. \(tmm), and use \(eom2), to obtain
$$
\ha\int_2dx^+d^2x_\perp T^{--}=
2\int_2dx^+d^2x_\perp \Bigl[(\del_+T^i)^2
	+ (\del_+\vp)\Bigl({1\over\del_\perp^2}\l\Bigr)\Bigr].
					\eqno(cont2)
$$
(We have again integrated by parts in the transverse directions
repeatedly.)  Now, the only subtle point concerns the first term on
the RHS of eq. \(cont2); we do not know the value of $T^i$ on the
boundary surfaces until the theory is solved.  In the free theory this
is no great difficulty---we can simply solve the theory and compute
this term (it is zero$^{\ddagger1}$\footnote{}{$^{\ddagger1}$
Specifically, with the solution of the free theory in hand we can
compute the integral over a finite-sized version of the surface shown
in Fig. 1, and afterwards let the size become infinite.  We find that
the contributions involving $T^i$ on the boundary surfaces
vanish.})---but in the interacting case it is crucial that the
generators be constructed only using the initial data.  However, $T^i$
must be integrable over the surface $x^+=0.$ This requires at a
minimum that $T^i$ should fall to zero for $x^-\ra\pm\infty.$ Note
that the field redefinition \(phi) insures that this is consistent;
the equation of motion for $T^i$ (eq. \(ateom)) does not contain any
terms that persist at $x^-=\pm\infty,$ so that if $T^i=0$ there
initially, it will remain so under evolution
in $x^+.$ This should be contrasted with the equation for the full
transverse gauge field $A^i$,
$$
(4\del_+\del_- - \del_\perp^2)A^i + \del_i\vp = 0.
					\eqno(eom3prime)
$$
Here it is {\it not} consistent to assume that $A^i$ vanishes on the
boundaries, unless $\vp=0$ for all $x^+.$

In a sense (which is necessarily somewhat imprecise in a continuum
formulation), the field redefinition \(atdefined) is a separation of
$A^i$ into a ``zero mode'' part and a ``normal mode'' part, with the
normal mode part assumed to vanish at large $|x^-|.$ (Recall that
$T^i$ is to be smeared in test functions that vanish at $k^+=0,$ so
that it contains no $x^-$-independent piece.)  This is reminiscent of
the situation in the discretized
formulation\r{my76,wittman88,hkw91b,mr92}, where the zero mode part of
any bosonic field is a constrained function of the other
fields in the theory, in this case $\vp.$  Examination of the equation
of motion \(eom3prime) shows that $A^i$ must contain an
$x^-$-independent piece equal to ${\del_i\over\del_\perp^2}\vp,$ or
this equation cannot be satisfied.

Thus we conclude that $T^i$ may be set to zero on the boundary
surfaces, giving the following expression for $P^-$:
$$
P^-=\ha\int_1dx^-d^2x_\perp (\del_iT^j)(\del_iT^j)
	+2\int_2dx^+d^2x_\perp
		(\del_+\vp)\Bigl({1\over\del_\perp^2}\l\Bigr).
					\eqno(finalpm)
$$
In a similar way we derive
$$
P^+=2\int_1dx^-d^2x_\perp (\del_-T^i)^2
	+ \ha\int_2dx^+d^2x_\perp (\del_iT^j)(\del_iT^j).
					\eqno(pplus)
$$
Again, the term involving $T^i$ on the boundary is zero, so that
$$
P^+=2\int_1dx^-d^2x_\perp (\del_-T^i)^2 .
					\eqno(finalpp)
$$

It is now a simple matter to deduce the field algebra, by demanding
that the Heisenberg relations reduce to the Euler-Lagrange equations.
(In general it may also be necessary to demand that the algebra of
generators be correctly realized, in order to fix all of the
commutators uniquely.)  The results are
$$
[T^i(x^-,x_\perp),\del^+T^j(y^-,y_\perp)]_{x^+=0}
=i\d^{ij}\d^{(3)}(\ub{x}-\ub{y})
					\eqno(tcommutator)
$$
$$
[\vp(x^+,x_\perp),\l(y^+,y_\perp)] = i
\d(x^+-y^+)\del_\perp^2\d^{(2)}(x_\perp-y_\perp)
					\eqno(plcommutator)
$$
$$
[A^i,\vp] = [A^i,\l] = [\vp,\vp] = [\l,\l] = 0 .
					\eqno(misccommutator)
$$
Thus the fields $\vp$ and ${1\over\del_\perp^2}\l$ are
canonically conjugate with respect to $x^-$-evolution.

Now let us solve the theory and compute the propagator.  We begin by
Fourier expanding the fields on their respective initial-value
surfaces:
$$
T^i(x^-,x_\perp)\Bigl|_{x^+=0} = {1\over(2\pi)^{3/2}} \int
{dk^+d^2k_\perp\over\sqrt{2k^+}}\Bigl(a_{\ub{k}i}e^{-i\ub{k}\cdot\ub{x}}
+a^\dagger_{\ub{k}i}e^{i\ub{k}\cdot\ub{x}}\Bigr)
					\eqno(tiinitial)
$$
$$
\vp(x^+,x_\perp) = {-i\over(2\pi)^{3/2}} \int {dp^-d^2p_\perp\over\sqrt2}
p_\perp^{1/2}\Bigl( f_{\ob{p}} e^{-{i\over2}p^-x^++ip_\perp x_\perp}
-f_{\ob{p}}^\dagger e^{{i\over2}p^-x^+-ip_\perp x_\perp}
\Bigr)
					\eqno(vpinitial)
$$
$$
\l(x^+,x_\perp) = {1\over(2\pi)^{3/2}} \int {dp^-d^2p_\perp\over\sqrt2}
p_\perp^{3/2}\Bigl( g_{\ob{p}}\ths e^{-{i\over2}p^-x^++ip_\perp x_\perp}
+g_{\ob{p}}^\dagger\ths e^{{i\over2}p^-x^+-ip_\perp x_\perp}
\Bigr),
					\eqno(linitial)
$$
where the factors of $p_\perp$ in \(vpinitial) and \(linitial) have
been introduced for later convenience.$^{\ddagger2}$
\footnote{}{$^{\ddagger2}$ We adopt
the notation $\ub{k}\equiv(k^+,k_\perp),$
$\ub{x}\equiv(x^-,x_\perp),$ and $\ub{k}\cdot\ub{x}\equiv\ha
k^+x^--k_\perp x_\perp.$ In addition, $\ob{p}=(p^-,p_\perp).$ In these
and subsequent formulas, integrals over $k^\pm$ are understood to run
from $0$ to $\infty,$ while integrals over transverse momenta run from
$-\infty$ to $\infty.$}
These last two expressions are already valid throughout
all of spacetime, since $\vp$ and $\l$ are independent of $x^-.$ The
solution for $T^i$ is obtained by inserting appropriate
$x^+$-dependent phases $e^{\pm{i\over2}k^-x^+}$ with $k^-$ given by
the free-particle dispersion relation
$$
k^-={k_\perp^2\over k^+}\ths .
					\eqno(freedr)
$$
Thus
$$
T^i(x^+,x^-,x_\perp) = {1\over(2\pi)^{3/2}}
\int{dk^+d^2k_\perp\over\sqrt{2k^+}}
\Bigl(
a_{\ub{k}i}
e^{-{i\over2}{k_\perp^2\over k^+}x^+-i\ub{k}\cdot\ub{x}}
+a^\dagger_{\ub{k}i}
e^{+{i\over2}{k_\perp^2\over k^+}x^++i\ub{k}\cdot\ub{x}}
\Bigr).
					\eqno(tieverywhere)
$$
The commutation relations \(tcommutator)--\(misccommutator) are
realized by the Fock space relations
$$
[a_{\ub{k}i} , a^\dagger_{\ub{l}j} ] = \d^{ij}
\d(k^+-l^+)\d^{(2)}(k_\perp - l_\perp)
					\eqno(acommutator)
$$
$$
[g_{\ob{p}},f^\dagger_{\ob{q}}] = [f_{\ob{p}},g^\dagger_{\ob{q}}]
=\d(p^+-q^+)\d^{(2)}(p_\perp - q_\perp)
					\eqno(fgcommutator)
$$
and
$$
[a,f]=[a,g]=[a^\dagger,f]=[a^\dagger,g]=[f,f^\dagger]=[g,g^\dagger]=0
					\eqno(mixed)
$$
If we wish, the mixed $f$ and $g$ commutators may be disentangled
through the redefinitions
$$
f_{\ob{p}}\equiv {1\over\sqrt{2}}(b_{\ob{p}}+ic_{\ob{p}})
\qquad\qquad g_{\ob{p}}\equiv {1\over\sqrt{2}}(b_{\ob{p}}-ic_{\ob{p}}),
					\eqno(fgredefs)
$$
where to reproduce \(fgcommutator)--\(mixed) we must have
$$
[a_{\ob{p}},a^\dagger_{\ob{q}}]=-[b_{\ob{p}},b^\dagger_{\ob{q}}]=
\d(p^--q^-)\d^{(2)}(p_\perp - q_\perp).
					\eqno(diagonalcomms)
$$
Thus we see clearly that the construction involves ghost states and an
indefinite-metric Hilbert space.  Unitarity and the full Poincar\'e
invariance will be recovered by restricting to a suitable physical
subspace, as discussed below.

We now have all the ingredients in hand to compute the propagator.  In
fact, the operator solution we have obtained (eqs. \(vpinitial),
\(linitial), and \(tieverywhere)) is identical to that obtained in
ref.\r{bdls85}, so we may be confident that the propagator will have
the ML form.  For completeness, however, we shall here discuss the
most singular component, $D^{--}(x).$ We have
$$
D^{--}(x)
=\vartheta(x^+)\langle0|A^-(x)A^-(0)|0\rangle
+\vartheta(-x^+)\langle0|A^-(0)A^-(x)|0\rangle ,
					\eqno(propagator)
$$
where $\vartheta(x)=1$ for $x>0$ and is zero otherwise.  We construct
the field $A^-$ following eq. \(aminus); it is the sum of a
``transverse'' part
$$
A^-_T(x) = -{\sqrt2\over(2\pi)^{3/2}}\sum_i\int dk^+d^2k_\perp
{k^i\over(k^+)^{3/2}}
\Bigl(a_{\ub{k}i}e^{-ik\cdot{x}}
+a^\dagger_{\ub{k}i}e^{ik\cdot{x}}\Bigr)
					\eqno(amtrans)
$$
and a ``longitudinal'' part
$$
\eqalignno{
A^-_L(x) &=
{\sqrt2\over(2\pi)^{3/2}}\int {dp^-d^2p_\perp\over p_\perp^{1/2}}
\Bigl[
\Bigl(g_{\ob{p}} - {p^-\over2p_\perp}f_{\ob{p}}\Bigr)
e^{-{i\over2}p^-x^++ip_\perp x_\perp} \cr
&\qquad\qquad\qquad +\Bigl(g_{\ob{p}}^\dagger-{p^-\over2p_\perp}
f_{\ob{p}}^\dagger\Bigr)
e^{+{i\over2}p^-x^+-ip_\perp x_\perp}
\Bigr]. &(amlong)
}
$$
{}From these, and the commutation relations \(acommutator)--\(mixed),
we easily compute, for example,
$$
\eqalignno{
\langle0|A^-(x)A^-(0)|0\rangle &=
{2\over(2\pi)^3}\int dk^+d^2k_\perp {k_\perp^2\over
(k^+)^3} e^{-ikx}\cr
&\qquad\qquad\qquad-
{2\over(2\pi)^3}\int dk^-d^2k_\perp {k^-\over k_\perp^2}
e^{-{i\over2}k^-x^++ik_\perp x_\perp}.&(vevmm)
}
$$
The expectation value of the fields in the other order is of course
just the complex conjugate of \(vevmm).  This is just the ML form of
$D^{--}$ (see for example ref.\r{bdls85}), written in terms of
light-cone variables.$^{\ddagger3}$
\footnote{}{$^{\ddagger3}$ Note that the choice of test functions in
which $T^i$ is to be smeared insures that the change from ordinary
spacetime variables to light-cone variables is well defined.}
The small-$k^+$ singularity in the first term in eq. \(vevmm) is
canceled by the large-$k^-$ region in the second term, as is easily
seen by making the change of variables $k^-={k_\perp^2\over k^+}$ in
the first term.

Finally, selection of the physical subspace follows exactly as
discussed in ref.\r{bdls85}.  The physical subspace is defined by the
requirement that Maxwell's equations and the Poincar\'e algebra are
obtained in matrix elements between physical states.  Now, the only
Maxwell equation that is not satisfied at the operator level is
\(eom2).  Therefore, physical states are those between which $\l$ has
vanishing matrix elements, or equivalently those satisfying
$$
g_{\ob{p}}|{\rm phys}\rangle = 0 .
					\eqno(physicality)
$$
It is straightforward to show that states defined in this way have
positive norm, so that unitarity holds in this subspace.  Furthermore,
one can check that the extra terms that occur in the algebra of the
Poincar\'e generators are all proportional to $\l$, so that the correct
Poincar\'e algebra is recovered in matrix elements between states
satisfying \(physicality).

\taghead{3.}
\subhead{3. Gauge Fields in Discretized Light-Cone Quantization}

In this section we shall remark briefly on the relevance of the above
when we require the gauge field to periodic in $x^-.$$^{\ddagger4}$
\footnote{}{$^{\ddagger4}$ We shall not consider the case where $A_\mu$ is
taken to be antiperiodic on $x^-,$ as the fermion bilinears we
eventually wish to couple to $A_\mu$ are necessarily periodic.} This
formalism is widely used in setting up actual numerical simulations of
light-cone field theories {\it \'a la} Tamm and Dancoff\r{bmpp93}.  It
is also of interest simply as a regulator of the small-$k^+$ region of
the theory.

Our first remark is by now well known: it is impossible to realize the
light cone gauge $A^+=0$ when the gauge field is required to be
periodic in $x^-.$  The easiest way to see this is to note that under
a gauge transformation
$$
A^+ \ra A^+ + \del^+\Lambda.
					\eqno(gaugetransf)
$$
In order to preserve the periodicity of the other components of
$A^\mu$ (or the boundary conditions imposed on some fermi field
coupled to $A_\mu$), the function $\Lambda$ must itself be
periodic.  Thus any $x^-$-independent part of $A^+$ is in fact gauge
invariant, and so an arbitrary $A^+$ cannot be brought to zero by a
gauge transformation.  The best we can do is to set everything
but the $x^-$-independent part of $A^+$ to zero; this is equivalent
to the gauge choice $\del_-A^+=0.$ The $k^+=0$ mode in $A^+$, which
can have nontrivial dependence on the transverse coordinates, must be
retained in the theory.

Another difference between the discretized and the continuum theories
is that in the discretized case it is natural, due to the explicit
imposition of boundary conditions on $A_\mu$, to eliminate all of the
residual gauge freedom.  The condition $\del_-A^+=0$ is preserved by
$x^-$-independent gauge transformations, so that one is free to impose
further gauge conditions that set various $x^-$-independent modes of
$A^-$ and $A^i$ to zero.  Kalloniatis and Pauli have proposed one
complete gauge fixing for discretized electrodynamics\r{kp93b}.  In
their construction the residual gauge freedom has been used to
eliminate a maximal number of components of the gauge field, so that
all that remain are physical fields, and various constrained zero
modes.  These latter objects occur generically in discretized theories
involving bosonic fields; they satisfy constraint relations, and must
be solved for in terms of the dynamical fields of the
theory\r{my76,wittman88,hkw91b,mr92}.

The result is a theory which is regulated at small $k^+,$ and which
involves only physical degrees of freedom.  As noted previously, in
the continuum theory we {\it could} fix boundary conditions at
$x^-=\pm\infty,$ thereby eliminating the residual gauge freedom.  The
resulting theory would not be isomorphic to that described in
ref.\r{bdls85}, of course, but would be a candidate formulation of
light-cone gauge field theory.$^{\ddagger5}$
\footnote{}{$^{\ddagger5}$ Note that for the light-cone gauge one
cannot eliminate the residual gauge freedom when quantizing on a
spacelike surface.} (The two formulations would be related by a gauge
transformation that rotates away the unphysical fields.)  All such
formulations that have so far been considered, however, give
inconsistent results in perturbative calculations, even for
gauge-invariant quantities.  In particular, we typically obtain the
principal value prescription for the spurious pole.

One lesson we take from this is that even though the occurrence of the
spurious pole is related to an unfixed gauge freedom, it matters what
we do to regulate it.  It is not true that it is simply a ``gauge
artifact,'' and any way of making it finite is acceptable.  This leads
us to ask whether anything analogous occurs in the discretized
formulation.  That is, does using discretization as a regulator result
in some inconsistency, analogous to using a principal value in the
continuum?  In the latter case, certain difficulties are apparent
already at the level of the free theory: we have positive (negative)
energy quanta propagating into the backward (forward) light cone.  It
will be important to check the viability of the discretized theory in
perturbative calculations, once a fully satisfactory formulation is
achieved.  This is a challenge because the relations that determine
the constrained zero modes are difficult to solve.  Work in this
direction has been recently reported by Kalloniatis and
Pauli\r{kp93b,kp93}.

One aspect of the discretized theory does have an analog in the
continuum theory described here.  Those modes of $A^i$ that are
constant in both $x^-$ and $x_\perp$ are the only ones that are not
either obviously dynamical (\ie, there is no classical equation of
motion for them) or determined by a constraint relation.  They do,
however, appear in \eg, the constraint relation for the constrained
part of the fermi field.  Physically, these degrees of freedom
correspond to photons with $k^+$ and $k_\perp$ vanishing, that is,
quanta propagating precisely along the surface $x^+=0.$ In the
continuum we can discard these by a suitable choice of test functions,
and feel reasonably confident that no serious damage is done because
the subset of the theory that is excluded is terrifically small.
Here, however, they are two of a countably infinite set of degrees of
freedom, and so may need to be included.  The point is that, precisely
because the quanta they represent propagate along $x^+=0,$ they should
be included in the theory as fields initialized along a surface of
constant $x^-,$ in the same way as $\vp$ and $\l.$ Work on including
these is currently in progress\r{mrinprep}.

\taghead{4.}
\subhead{4. Discussion}

We have seen that the results of ref.\r{bdls85} can be reproduced in
light-cone quantization if proper attention is paid to the degrees of
freedom and where they must be initialized.  In particular, the
unphysical fields $\vp$ and $\l$ must be initialized on a surface of
equal $x^-$, and satisfy equal-$x^-$ canonical commutation relations.
A further complication is the necessity of including boundary
contributions in the calculation of conserved charges.  This becomes
especially tedious in the interacting theory; the fully interacting
case, with application to light-cone bound state equations, will be
discussed elsewhere\r{mr93b}.

In this language, it is the appearance of the second characteristic
surface which brings in the ``dual'' gauge vector $n^*_\mu=(1,0,0,-1)$
needed to formulate the ML prescription\r{bassettobook}.  It is also
clear why the ML prescription is impossible to realize in the context
of strictly $x^+$-ordered (light-cone) perturbation theory---we simply
do not have a formalism in which all fields are evolved from a surface
of constant $x^+.$ This is always a possibility when quantizing on
characteristic surfaces, though in certain cases it may be avoided.
The discretized formulation is not precisely analogous, in part
because we cannot realize the light-cone gauge for periodic gauge
fields.  It does appear to have some similar features, however, in
particular the need to include fields that live along other
characteristics.  It will be interesting to investigate whether
discretized light-cone QED suffers from pathologies analogous to those
that seem to plague continuum formulations in the light cone gauge,
when the gauge is completely fixed.

Finally, we should perhaps emphasize that this entire discussion has
had as its motivation the successful formulation of perturbation
theory for gauge theories.  The relevance of the ML prescription for
the kind of {\it nonperturbative} calculations everyone would like to
do is somewhat unclear.  Perhaps these problems only arise if
we insist on setting up a calculational scheme in terms of unphysical,
gauge-dependent quantities, like propagators and vertices.  It may
very well be that we have to be {\it more} careful in this case than
we would have to be if we we formulated and solved the theory in a
physical way from the start.

\subhead{ACKNOWLEDGMENTS}

\noindent
It is a pleasure to thank K. Hornbostel and A. Kalloniatis for many
helpful discussions.  This work was supported in part by the U.S.
Department of Energy under grant no. DE-FG05-92ER40722.  DGR also thanks
the Lightner-Sams Foundation for its generous support.

\vfill\eject
\references

\refis{bassettobook}A. Bassetto, G. Nardelli, and R. Soldati,
	{\it Yang-Mills Theories in Algebraic Non-Covariant Gauges}
	(World Scientific, 1991)

\refis{bassetto91}A. Bassetto, {\it The Light-Cone Feynman Rules
	Beyond Tree Level,} DFPD-91-TH-19  (Invited talk given at the 14th
	Workshop on Problems in High-Energy Physics and Field Theory,
	Protvino, U.S.S.R., 1991)

\refis{bdls85}A. Bassetto, M. Dalbosco, I. Lazzizzera, and R. Soldati,
	\pr, D31, 1985, 2012,

\refis{bds87}A. Bassetto, M. Dalbosco, and R. Soldati, \pr, D36, 1987, 3138,

\refis{bmpp93}S.J. Brodsky, G. McCartor, H.-C. Pauli, and S. Pinsky,
	{\sl Particle World \bf 3 \rm (1993) 109}

\refis{casher76}A. Casher, \pr, D14, 1976, 452,

\refis{hkw91b}T. Heinzl, S. Krusche, and E. Werner, \pl, 272B, 1991, 54,

\refis{kp93}A.C. Kalloniatis and H.-C. Pauli, Heidelberg preprint MPIH-V6-1993

\refis{kp93b}A.C. Kalloniatis and H.-C. Pauli, Heidelberg preprint
MPIH-V15-1993

\refis{ks70}J.B. Kogut and D.E. Soper, \pr, D1, 1970, 2901,

\refis{leibbrandt84}G. Leibbrandt, \pr, D29, 1984, 1699,

\refis{ls93}H.H. Lu and D.E. Soper, \pr, D48, 1993, 1841,

\refis{mandelstam83}S. Mandelstam, \np, B213, 1983, 149,

\refis{my76}T. Maskawa and K. Yamawaki, \ptp, 56, 1976, 270,

\refis{mccartor88}G. McCartor, \zphys, C41, 1988, 271,

\refis{mr92}G. McCartor and D.G. Robertson, \zphys, C53, 1992, 679,

\refis{mr93b}G. McCartor and D.G. Robertson, {\it Light-Cone Quantization
	of Gauge Fields: II. Bound State Equations in a Tamm-Dancoff
	Truncation,} in preparation

\refis{mrinprep}G. McCartor and D.G. Robertson, in preparation

\refis{ss72}S. Schlieder and E. Seiler, \cmp, 25, 1972, 62,

\refis{soperheidelberg}D.E. Soper, talk presented at the Workshop on Gauge
	Field Theory on the Light Cone, Heidelberg, June 1991

\refis{tomboulis73}E. Tomboulis, \pr, D8, 1973, 2736,

\refis{wittman88}R.S. Wittman, in {\it Nuclear and Particle
	Physics on the Light Cone,} M.B. Johnson and L.S. Kisslinger,
	eds. (World Scientific, 1988)

\endreferences
\endit
\end